\documentclass
[preprint,twocolumn,10pt,nofootinbib,amsfonts,amssymb,bibnotes,pre]{revtex4}%
\usepackage{amsfonts}
\usepackage{amsmath}
\usepackage{amssymb}
\usepackage{graphicx}%
\begin{document}
\preprint{UATP/07-01}
\title{Entropy Crisis, Defects and the Role of Competition in Monatomic Glass Formers}
\author{P. D. Gujrati}
\email{pdg@arjun.physics.uakron.edu}
\affiliation{Departments of Physics and of Polymer Science, The University of Akron, Akron,
OH 44325}
\date{\today}

\begin{abstract}
The demonstration here of an entropy crisis in monatomic glass formers along
with previously known results finally establishes that the entropy crisis is
ubiqutous in all supercooled liquids. We show that interactions that compete
with crystal order weakens the ability to form glass. We also study the
defects in the ideal glass; they are very different from those in the crystal.

\end{abstract}
\maketitle

A common feature of almost all materials is that they can become glassy, i.e.
an amorphous solid, representing a \emph{metastable} state \cite{Kauzmann}.
The glass is obtained by cooling the supercooled liquid (SCL), and need not
obey the third law of thermodynamics, which is only valid for equilibrium
states \cite{Landau}. The true equilibrium state corresponds to a crystalline
state. The configurational entropy of SCL exhibits a rapid drop below the
melting temperature $T_{\text{M}}$ \cite{Kauzmann}, and eventually vanishes at
some temperature $T=$ $T_{\text{K}}<T_{\text{M}}.$ It will become
\emph{negative if extrapolated} to lower temperatures. Since a negative
entropy is \emph{unphysical}, an ideal glass transition must intervene to
avoid this \emph{entropy crisis} at $T_{\text{K}}$. In the limit of zero
cooling rate (not accessible in experiments or simulations, but accessible in
a theoretical setup) in the metastable region, metastable states become
\emph{stationary}, and can be described by "equilibrium" statistical mechanics
by \emph{restricting} the allowed microstates to be disordered; microstates
leading to the crystalline state are \emph{not} allowed. In experiments, the
extrapolated configurational entropy of many glassy materials at $T=0$
\cite{GujGold0} is found to have a non-zero value depending on the rate of
cooling. This does not rule out the possibility that the entropy of the
hypothetical "stationary" glass vanishes in the limit of zero cooling rate at
a non-zero temperature at a positive temperature.

Demonstrating an entropy crisis for supercooled liquids in a restricted
formalism has been one of the most challenging problems in theoretical
physics. An entropy crisis in long polymers was theoretically demonstrated
almost fifty years ago by Gibbs and DiMarzio \cite{GibbsDiMarzio} to support
the entropy crisis as a \emph{fundamental principle} underlying glass
transitions in long polymers. Their work was later severely criticized by
Gujrati and Goldstein \cite{GujGold} for its poor approximation, and doubt was
cast on whether the entropy crisis in long polymers was genuine. The situation
changed when in a recent work, Gujrati and Corsi \cite{GujC} established the
existence of an entropy crisis in long polymers by using a highly reliable
approximate approach. This was very important as the idea of Gibbs and Di
Marzio has been pivotal in shaping our thinking about the ideal glass
transition. Recently, we have also succeeded in demonstrating the entropy
crisis in a dimer model \cite{GujS} containing anisotropic interactions.
However, to establish the entropy crisis as a fundamental principle
underliving the glass transition in \emph{all} supercooled liquids, we need to
demonstrate the crisis in simple isotropic fluids containing monatomic
particles. So far, this has not been feasible in any theoretical approach and
is one of the outstanding theoretical physics problems for a complete
understanding of glass transition.

Here, we establish an entropy crisis in monatomic systems, thereby finally
succeeding in establishing the entropy crisis as a fundamental mechanism
driving glass transitions in all supercooled liquids. Study of monatomic glass
formers will allow us to obtain a better understanding of the glassy structure
(defects therein) whose accurate representation remains still challenging. The
slow relaxation \cite{Kauzmann} in SCLs is similar to that observed in
ordinary spin glasses \cite{Morgenstern}. Therefore, it is not surprising that
ordinary monatomic glasses have been traditionally modeled as spin glasses,
whose important features are their geometrical frustration and competition. It
is commonly believed that the competition and frustration play an important
role in promoting the glassy behavior. We will also check this hypothesis in
this work.

It should be noted that frustrated antiferromagnets (AF) and spin glasses do
not usually possess long range order at low temperatures because of \ a highly
degenerate ground state \cite{Ramirez} and their glassy behavior is brought
about by the presence of frustration or quenched impurities and is somewhat
well understood. In contrast, supercooled liquids require a \emph{unique}
ground state, the crystal. This distinguishes the glassy behavior in
supercooled liquids and requires considering an \emph{unfrustrated} AF model
as a paradigm of simple fluids or alloys. We consider a pure \ (no frustration
or quenched impurities) AF Ising model, which possesses a unique ordered state
so that supercooling can occur. This then results in a glassy state. We are
not aware of any simple model calculation to date to justify glassy states in
a pure AF model. We also find that the competition considered in this work
inhibits instead of promoting the glass transition, which is a surprising result.

\textbf{Model. }We introduce\ the following \emph{AF} \emph{Ising model} in
zero magnetic field on a square or a cubic lattice (lattice spacing $a$) with
the interaction energy%
\begin{equation}
E\mathcal{=}J\sum SS^{\prime}+J^{\prime}\sum SS^{\prime}S^{\prime\prime
}.\label{BinaryE}%
\end{equation}
The first sum is over nearest-neighbor spin pairs and the second over
neighboring spin triplets, which we take to be within a square for simplicity.
We take $S=\pm1$ to denote A, and B particles for an alloy, or the particle
and void for a fluid. For $\left\vert J^{\prime}\right\vert \leq2J$, we have
an AF ordering\ at low temperatures with a sublattice structure:\ spins of a
given orientation are found preferentially on one of the two sublattices.
Antiferromagnetically ordered squares (AFS) with spins alternating, and
ferromagnetically ordered squares (FS) with spins the same are the only two
square conformations that contribute to the first term in (\ref{BinaryE}). We
may identify the AF ordered structure as a crystal \cite{GujS}. For a fluid,
this model represents a strong repulsion at a lattice spacing $a$, and
attraction at lattice spacings $\sqrt{2}a$ between particles. For $\left\vert
J^{\prime}\right\vert \geq2J,$ the AF ordering is destroyed at low
temperatures; $S$ is the same everywhere. We set $J$=1 to set the temperature
scale and only consider $\left\vert J^{\prime}\right\vert \leq2J$. It is easy
to see that the free energy depends on $\left\vert J^{\prime}\right\vert ,$
not on\ its sign. In particular, the ground state energy per spin of the AF
ordered state is $E_{0}=-2J,$ regardless of $J^{\prime}.$ In the following, we
will measure the energy and the free energy with respect to the ground state
to give the excitation energies. In this case, both will vanish at $T=0.$ The
non-zero value of $\left\vert J^{\prime}\right\vert $\ creates a preference
for the product $SS^{\prime}S^{\prime\prime}$ in a square to be of a fixed
\ sign, which then competes with the formation of the crystal in which this
product can be of either sign. A positive $\ ($negative$)$ $J^{\prime}%
$\ provides a preference for $S=-1$ ($+1$), so $J^{\prime}$ can be used to
also control the abundance of one of the spin states.

The entropy $S(T)$ of the model cannot be negative if the state has to occur
in Nature or simulations; indeed, neither can ever show any entropy crisis. If
the metastable state entropy $S(T)=0$ at a positive temperature $T_{\text{K}%
},$ then its \emph{extension} will experience an entropy crisis and must be
replaced by an \emph{ideal glass} below $T_{\text{K}},$ the \emph{ideal glass
transition temperature }\cite{note0}. The model cannot be solved exactly
except in one dimension. It is usually studied in the mean-field approximation
commonly known as the Bragg-Williams approximation \cite{Kubo} adapted for an
AF case. However, the approximations is known to be very crude. Indeed, Netz
and Berker \cite{Berker} have shown that one of the shortcomings of the
approximation is that it abandons the hard-spin condition $S^{2}=1.$ This
condition is easily incorporated in exact calculations on recursive lattices
\cite{GujPRL} and it was discovered that such calculations are more reliable
than the conventional mean-field approximations. Therefore, we adopt the
recursive lattice approach here.

We consider a homogeneous Husimi cactus in which $q$ squares meet at a site.
The model is solved exactly on the cactus as described in \cite{GujPRL}. The
cactus can be thought as an approximation of a square lattice for $q=2$ or a
cubic lattice for $q=3$, so that the exact Husimi cactus solution can be
thought of as their approximate solution.

\textbf{Solution. }We follow \cite{GujPRL} and solve the model recursively. We
label sites on the cactus by an index $m$, which increases\ sequentially
outwards from $m=0$ at the origin. We introduce partial PF's $Z_{m}(\uparrow)$
and $Z_{m}(\downarrow),$ depending on the states of the spin at the $m$-th
cactus level. It represents the contribution of the part of the cactus above
that level to the total PF. We then introduce the ratio
\begin{equation}
x_{m}\equiv Z_{m}(\uparrow)/[Z_{m}(\uparrow)+Z_{m}(\downarrow)],
\label{Ratios}%
\end{equation}
which satisfies the recursion relation (RR)
\begin{equation}
x_{m}\equiv\frac{f(x_{m+1},x_{m+2},v)}{f(x_{m+1},x_{m+2},v)+f(y_{m+1}%
,y_{m+2},1/v)}, \label{RR}%
\end{equation}
where $f(x,x^{\prime},v)\equiv x^{2r}x^{\prime r}/uv^{2}+2x^{r}x^{\prime
r}y^{r}v+x^{2r}y^{\prime r}v+ux^{\prime r}y^{2r}+2x^{r}y^{r}y^{\prime
r}+y^{2r}y^{\prime r}/v$ with $r=q-1$\ and where $u\equiv e^{4\beta},v\equiv
e^{2\beta J^{\prime}},y\equiv1-x,y^{\prime}\equiv1-x^{\prime}.$

There are two kinds of fix-point (FP) solutions of the RR (\ref{RR}) that
describe the bulk behavior \cite{GujPRL}. In the 1-cycle solution, the FP
solution becomes independent of the level $m$, and is represented by $x^{\ast
}.$ For $J^{\prime}=0$, $x^{\ast}$ is given by $x^{\ast}=1/2$, as can be
checked explicitly. For $J^{\prime}\neq0,$ $x^{\ast}\neq1/2$ and has to be
obtained numerically. This solution \emph{exists} at all temperatures
$T\geq0;$ thus, there is no spinodal of this solution$.$ This solution
describes the disordered phase. At $T\rightarrow\infty,$ all spins are
uncorrelated so that the density per site $\phi_{\text{FS}}=1/8,$ and
$\phi_{\text{AFS}}=1/16,$ and the entropy is $S=\ln2.$ As $T$ is reduced,
$\phi_{\text{FS}}$ decreases, while $\phi_{\text{AFS}}$ increases. The other
FP solution of interest is a 2-cycle solution associated with the AF state
containg AFSs \cite{GujPRL}. It alternates between two values $x_{1}^{\ast},$
and $x_{2}^{\ast}$ which occur at successive levels. This kind of FP solution
has also been observed in other systems such as semi-flexible polymers
\cite{GujC,GujRC}, dimers \cite{FedorDimer}$,$ and stars and dendrimers
\cite{CorsiThesis}, and has been thoroughly investigated. At $T=0,$ the
2-cycle solution is given by $x_{1}^{\ast},x_{2}^{\ast}=1,0$ or $0,1$
describing the perfect crystal ($\phi_{\text{FS}}=0$,$\phi_{\text{AFS}}%
=0.5$)$.$ This solution then evolves with $T$ due to excitations and describes
the crystal at low temperatures. The free energy is calculated by the general
method due to Gujrati \cite{GujPRL,GujC,CorsiThesis}. Whichever solution has
the lower free energy represents the equilibrium state. The solution with the
higher free energy, then, represents the metastable state, which can only be
observed in Nature if its entropy remains non-negative. The temperature where
the two solutions have the same free energy is the transition temperature,
which we denote by $T_{\text{M}}.$%
\begin{figure}
[ptb]
\begin{center}
\includegraphics[
trim=1.204846in 3.885781in 1.469097in 3.107746in,
height=1.9865in,
width=2.8807in
]%
{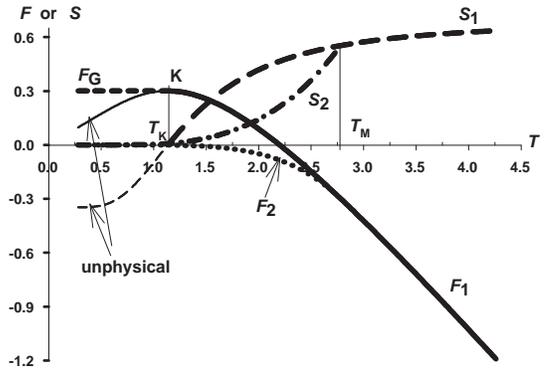}%
\caption{The free energy and entropy for the two FP solutions. The model shows
an entropy crisis and an ideal glass transition at $T_{\text{K}}.$ The thin
curves represent unphysical states (negative entropy) and are replaced by the
ideal glass state.}%
\label{F1}%
\end{center}
\end{figure}

\textbf{Results. }The results for $q=2$ are presented in Figs. 1-3. The free
energy $F_{1}$ and entropy $S_{1}$ associated with the 1-cycle FP solution are
shown by the continuous and the long dash curves in Fig. 1. The free energy
$F_{2}$ and entropy $S_{2}$ associate with the 2-cycle FP solution are shown
by the dotted and the dash-dot curves. The energy $E(T)$ as a function of $T$
and the entropy $S(E)$ as a function of $E$ are shown in Fig. 2. We have set
$J^{\prime}=0.01$ for figures 1, and 2. As said earlier, $F$ and $E$ represent
the contributions of excitations with respect to the ground state energy
$E_{0}=-2J$, so that they vanish at $T=0,$ as is clearly seen in Figs. 1, and
2. The transition temperature is found to be $T_{\text{M}}\cong2.7706$. We see
from Fig. 1 that $F_{1\text{ }}$crosses zero and becomes positive below
$T=T_{\text{eq}}\simeq2.200$ but again becomes zero (not shown here, but we
have checked it) as $T\rightarrow0.$ Thus, $F_{1}$ possesses a maximum at an
intermediate temperature (see point K in Fig. 1) at $T=T_{\text{K}}%
\simeq1.1316,$ so that the entropy $S_{1}$ vanishes there. Below $T_{\text{K}%
}$, the \emph{continuation} of $F_{1}$ and $S_{1}$, shown by their thin
portions in Figs. 1 and 2, continue to satisfy the \emph{stability condition}
(non-negative specific heat). Despite this, they \emph{cannot} represent any
physical states in the system due to \emph{negative entropy} and have to be
discarded as \emph{unphysical}. Below $T_{\text{K}}$, we must extend the
metastable state (described by $F_{1}$ and $S_{1}$ between $T_{\text{K}}$ and
$T_{\text{M}})$ by a glassy phase of a constant free energy $F=F_{\text{G}},$
see the short dash horizontal line in Fig. 1, and $S=S_{\text{G}}=0.$ The
1-cycle energy at K is $E_{1\text{K}}=F_{\text{G}}\simeq0.301.$ The entropy
$S_{2}$ is never negative, and the 2-cycle FP solution represents the
equilibrium crystal below $T_{\text{M}}.$%
\begin{figure}
[ptb]
\begin{center}
\includegraphics[
trim=0.603722in 3.086035in 1.676735in 2.595462in,
height=2.4076in,
width=2.8435in
]%
{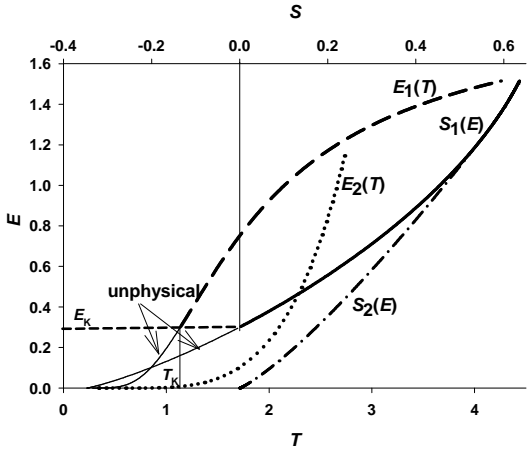}%
\caption{$S-E-T$ relationship for the two FP solutions. The excitations in the
two near $T=0$ are very different. The excitations in the 1-cycle state near
$T_{\text{K}}$ are strongly interacting as opposed to those near $T=0.$}%
\label{F2}%
\end{center}
\end{figure}

\textbf{Competition with Crystal Ordering. }The behavior as a function of
$\left\vert J^{\prime}\right\vert $ of the transition temperature
$T_{\text{M}}$ (empty circles), the ideal glass transition temperature
$T_{\text{K}}$ (filled circles) and their ratio $T_{\text{M}}$/$T_{\text{K}}$
(triangles) are shown in Fig. 3. As said above, $\left\vert J^{\prime
}\right\vert $ competes with the crystal ordering and reduces $T_{\text{M}}.$
One can use the inverse ratio $T_{\text{K}}$/$T_{\text{M}}$ as a
\emph{measure} of the relative ease of glass formation: larger this value,
easier it is to obtain the ideal glass as $T_{\text{K}}$ is not too deep
relative to $T_{\text{M}}.$ What we observe is that $T_{\text{M}}%
$/$T_{\text{K}}$ increases with $\left\vert J^{\prime}\right\vert ,$ with
$T_{\text{K}}$ approaching zero faster than $T_{\text{M}},$ so that the ratio
$T_{\text{M}}$/$T_{\text{K}}$ continues to increase with $\left\vert
J^{\prime}\right\vert .$ This implies that it becomes harder to obtain the
ideal glass as $T_{\text{K}}$ becomes relatively farther away from
$T_{\text{M}}$ as $\left\vert J^{\prime}\right\vert $ increases$.$ The
competition provided by $\left\vert J^{\prime}\right\vert $ weakens not only
crystal ordering but also "weakens" forming the ideal glass. Consequently,
competition does not enhance the ability to undergo ideal glass transition, an
interesting result which is being explored further \cite{Cerena} to see if
other competitions behave similarly.

\textbf{Analysis of Defects. }To understand the difference between the
disordered liquid and the crystal defects, we turn to Fig. 2 and observe that
near $T=0,$ the excitation energies of both FP solutions are very different,
even though $E_{1}(0)=E_{2}(0)=0$. Detailed analysis will be pesented
elsewhere \cite{Keith}. The excitations (defects) in the crystal are known to
be due to \emph{point-like} excitations caused by the reversal of a single
spin which changes the free energy by $\simeq8J$ (coordination number 4) with
respect to the ground state$;$ here we assume $J^{\prime}$ to be small.
Therefore, this excitation causes the leading term in the free energy $F_{2}$
to be $1/u^{2}$ \cite{Domb} and can be treated as \emph{non-interacting} as
long as they are small in number. What kinds of excitations are deducible from
the form of $F_{1}$ near $T=0?$ To answer this, we consider the simple case of
$J^{\prime}=0,$ which also gives rise to $F_{1}$ over the entire temperature
range $T\geq0.$ Here, we can carry out an analytical investigation since
$x^{\ast}=1/2.$ As we are only interested in the excitation energy, we will
overlook the unphysical nature of the entropy near $T=0$ to study the
excitation spectrum. The excitation energy [=4($\phi_{\text{FS}}%
-\phi_{\text{AFS}}$)+2] and free energy are given by ($w=3+u/2+1/2u$)%
\[
E_{1}(T)=-(u-1/u)/w+2,F_{1}(T)=-(T/2)\ln w+2,
\]
where $\phi_{\text{FS}}=1/4uw,$ and $\phi_{\text{AFS}}=u/4w.$\ The entropy is
calculated using $S_{1}(T)=(E_{1}-F_{1})/T.$ At $T=0,E_{1}(0)=0$ implies that
all squares are. Near $T=0,$ we find that $E_{1}(T)\simeq12/u,$~$F_{1}%
(T)\simeq T(%
\frac12
\ln2-3/u)$ and $S_{1}(T)\simeq-%
\frac12
\ln2+3/u+12\beta/u.$ ($S_{1}$ at $T=0$ in Fig. 1 is almost $-%
\frac12
\ln2$.) What we discover is that the excitations due to $1/u$-term near $T=0$
are not the \emph{uncorrelated} single spin reversal in the background of a
perfect crystal. Rather, they represent \emph{correlated} excitations in the
form of FS in the background of AFSs by turning an AFS into a FS. Each FS
excitation requires an energy $4J$ per site. This is in accordance with the
general discussion above. Indeed, the excitation spectrum is given by the
expansion of $E_{1}(T)$ in powers of $1/u$ \cite{Domb}. These excitations also
explain why the thin portion of $E_{1}(T)$ rises more rapidly than $E_{2}(T)$
in Fig. 2 so that $E_{\text{K}}\simeq0.3$ is appreciably higher than the
$E_{2}(T)\simeq0.1$ at $T_{\text{K}}$ due to a lower $\phi_{\text{AFS}}%
\simeq0.426$ and higher $\phi_{\text{FS}}\simeq3.58\times10^{-4}$ in the
metastable state. In contrast, $\phi_{\text{AFS}}\simeq0.498,$ and
$\phi_{\text{FS}}\simeq7.991\times10^{-7}$ in the crystal. This is consistent
with most defects in the crystal being point defects. As there is no
non-analyticty at $T_{\text{K}}$ in SCL$,$ the excitation spectrum remains
continuous above $T_{\text{K}}$ where physical states occur. The density of
sites $\phi_{\text{unc}}$ not covered by AFSs and FSs is $\phi_{\text{unc}%
}\simeq0.148$ in SCL and $\phi_{\text{unc}}\simeq3.684\times10^{-3}$ in the
crystal$.$ These sites are probably uncorrelated in SCL, but this needs to be
carefully checked. Nevertheless, the above analysis leads us to conclude that
the excitations at or above $T_{\text{K}}$ in SCL are very different from the
point defects of the crystal and are the ones that get frozen in the glass
that is formed at $T_{\text{K}}.$ For $T<T_{\text{K}},$ we have an ideal
glass, shown by the horizontal short dash curve in Fig. 2, of \ constant
energy $E_{\text{K}}$ and zero entropy.%
\begin{figure}
[ptb]
\begin{center}
\includegraphics[
trim=1.023864in 3.497862in 1.471646in 3.023129in,
height=2.0185in,
width=2.6749in
]%
{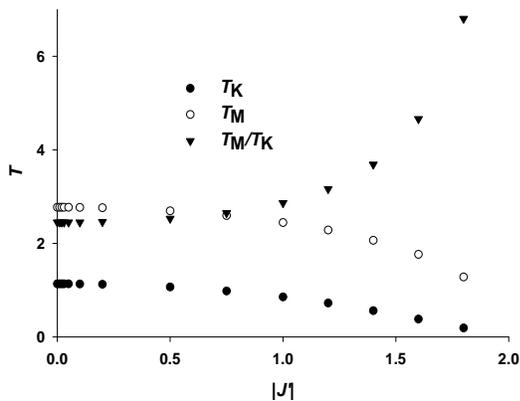}%
\caption{The effect of $\left\vert J^{\prime}\right\vert ,$ which creates
competition with the crystal ordering, on $T_{\text{M}},$ and $T_{\text{K}}$
and their ratio. The weakening of crystal ordering also "weakens" the ideal
glass formation. }%
\label{F3}%
\end{center}
\end{figure}

The excitation spectra of both solutions over the physical range are
completely described by their respective entropies $S_{1}(E)$ (the thick part)
and $S_{2}(E)$ shown in Fig.2. For $E<E_{\text{K}},$ the excitations in the
glass cannot change since they are frozen at constant energy $E_{\text{K}},$
but continue to change in the crystal. In experiments, the ideal glass will
never be observed due to time-limitations and one would obtain a
non-stationary state whose entropy $\widetilde{S}_{1}(E)$ must satisfy
$\widetilde{S}_{1}(E)\leq$ $S_{1}(E)$ according to the law of increase of
entropy \cite{Landau}$.$ In this case, the non-stationary glass will have some
excitations at low temperatures.

For the AF case that we consider here, the 1-cycle solution is found to exist
at \emph{all} temperatures and describes the disordered liquid above and its
metastable continuation below the transition temperature. There is no
singularity in this fix-point solution at the transition. In contrast, for the
ferromagnetic case ($J<0$), the 2-cycle FP solution is never stable, and the
1-cycle solution has a singularity at the ferromagnetic transition and its
entropy never becomes negative.

We have already shown elsewhere \cite{FedorDimer} that the glassy state for
dimers contains a higher density of voids than the corresponding crystal at
the same temperature. This is also true in the current model. The voids
distribute themselves in the lattice at equilibrium, and the corresponding
1-cycle solution gives rise to an excitation spectrum so that $S_{1}(E)$
vanishes at $T_{\text{K}}.$ If it happens that the system is quenched, then
all we can say is that the corresponding spectrum $\widetilde{S}_{1}(E)$ of
the quenched system must satisfy the standard condition $\widetilde{S}%
_{1}(E)\leq$ $S_{1}(E)$. Despite this, it is possible that the entropy of the
quenched system does not vanish at a positive temperature. There is no contradiction.

To summarize, our model calculation demonstrates that monatomic systems also
give rise to an ideal glass, thereby making the entropy crisis ubiquitous. The
glass contains correlated defects that are very different from those in the
crystal. Moreover, the competition does not enhance the ability to form a
glass. We have only investigated a classical model and it would be interesting
to see if quantum calculations support this picture.


\begin{thebibliography}{99}
\bibitem {note0}Even though no experiment can ever detect the ideal glass
transition due to time constraints, its existence most certainly implies that
real glasses would be formed during experiments.
\end{thebibliography}
\end{document}